\documentclass[12pt,preprint]{aastex}
\begin{document}

\parindent=1.0cm

\baselineskip=0.8cm

\title {The Young Open Clusters King 12, NGC 7788, and NGC 7790: Pre-Main Sequence Stars and Extended Stellar Halos\altaffilmark{1}}

\author{T. J. Davidge}

\affil{Dominion Astrophysical Observatory,
\\National Research Council of Canada, 5071 West Saanich Road,
\\Victoria, BC Canada V9E 2E7}

\altaffiltext{1}{Based on observations obtained with MegaPrime/MegaCam, a joint
project of CFHT and CEA/DAPNIA, at the Canada-France-Hawaii Telescope (CFHT)
which is operated by the National Research Council (NRC) of Canada, the
Institut National des Sciences de l'Univers of the Centre National de la
Recherche Scientifique (CNRS) of France, and the University of Hawaii.}

\begin{abstract}

	The stellar contents of the open clusters King 12, NGC 7788, and NGC 
7790 are investigated using MegaCam images. Comparisons with isochrones yield an age 
$< 20$ Myr for King 12, 20--40 Myr for NGC 7788, and 60 -- 80 Myr for NGC 7790 based 
on the properties of stars near the main sequence turn-off (MSTO) in each cluster. 
The reddening of NGC 7788 is much larger than previously estimated. The 
luminosity functions (LFs) of King 12 and NGC 7788 show breaks that are 
attributed to the onset of pre-main sequence (PMS) objects, 
and comparisons with models of PMS evolution yield ages that 
are consistent with those measured from stars near the MSTO. 
In contrast, the $r'$ LF of main sequence stars in NGC 7790 
is matched to $r' = 20$ by a model that is based on the 
solar neighborhood mass function. The structural properties of all three 
clusters are investigated by examining the two-point angular correlation 
function of blue main sequence stars. King 12 and NGC 7788 are 
each surrounded by a stellar halo that extends out to 5 
arcmin ($\sim 3.4$ parsecs) radius. It is suggested that these halos form 
in response to large-scale mass ejection early in the evolution of the 
clusters, as predicted by models. In contrast, blue main sequence stars 
in NGC 7790 are traced out to a radius of $\sim 7.5$ arcmin ($\sim 5.5$ parsecs), 
with no evidence of a halo. It is suggested that all three clusters may 
have originated in the same star-forming complex, but not in the same 
giant molecular cloud.

\end{abstract}


\section{INTRODUCTION}

	Star clusters are recognized as key probes for tracing the 
star-forming history (SFH) and chemical evolution of galaxy disks. However, 
the substantial numbers of young field stars in the Galaxy 
(e.g. Lada \& Lada 2003; Bonatto \& Bica 2011a) and in nearby galaxies (e.g. 
Silva-Villa \& Larsen 2011) indicate that most clusters are short-lived, 
and this should be considered when interpreting cluster surveys. 
Those clusters that survive their early evolution likely do so because of a 
compbination of initial conditions and the timing of gas ejection (e.g. Smith et 
al. 2011). Even massive clusters that survive their early evolution may not 
remain intact for more than a few disk crossing times. For example, a 10$^4$ 
M$_{\odot}$ cluster -- which is substantially more massive 
than the majority of clusters in the solar neighborhood -- may survive for 
no more than 1.3 Gyr in the relatively benign environment of the Galactic disk 
(e.g. Lamers et al. 2005).

	Once free of their natal clusters, stars can disperse over large 
distances on cosmologically short timescales. Bastian et al. (2011) conclude that 
the timescale for stars to diffuse over galaxy-wide scales in nearby dwarf irregular 
galaxies is $\sim 100$ Myr. Davidge et al. (2011) investigate the clustering 
properties of stars in the outer disk of M33, and conclude that cluster stars 
in this environment may disperse over kpc scales on timescales of tens of Myr.

	The study of star clusters with ages of a few tens of Myr will 
provide clues into the dispersal of stars from their birth environments. In addition, 
as demonstrated by Bonatto \& Bica (2011b), studies of young star clusters that are 
physically close together, and so may have shared a common birth place, will 
provide constraints on the SFHs and lifetimes of Giant Molecular Clouds (GMCs). 
In the present study, moderately deep $u*, g',$ and $r'$ images obtained 
with MegaCam on the Canada-France-Hawaii Telescope (CFHT) are used to 
examine the stellar contents of the young clusters King 12, NGC 7788, and NGC 7790. 
These clusters are separated on the sky by only a few tens of arc minutes, and so 
can be observed with a single MegaCam pointing. That these clusters are close 
together on the sky (projected separation $\sim 20$ parsecs)
and have ages that agree to within a few tens of Myr, opens the possibility that they may 
have originated in the same star-forming complex, and possibly even the same GMC.

	Various properties of the clusters, taken from the WEDBA database (Mermilliod 
1995), are summarized in Table 1. King 12 and NGC 7788 do not have published 
color-magnitude diagrams (CMDs) that are based on data obtained with modern detectors. 
The sole photometric study of King 12 is that of Mohan \& Pandey (1984), who present 
photoelectric photometry of 30 stars with $V \leq 14$, and rank King 12 between NGC 
2362 and NGC 884 in terms of age. The only published CMD of NGC 7788 was produced by 
Becker (1965) from photographic observations. He measures 
$E(B-V) = 0.28$ for this cluster, which is markedly lower than the 
reddenings of King 12 and NGC 7790.

	NGC 7790 has been the subject of various studies, motivated in large part by the 
presence of 3 Cepheids, making this cluster an important rung in the cosmic 
distance ladder. Deep imaging of NGC 7790 suggests that $E(B-V) = 0.54$, with a distance 
modulus of $12.65 \pm 0.15$ and an age in the range 50 -- 100 Myr (Romeo et 
al. 1989). Matthews et al. (1995) investigate the photometric and radial velocity 
variations of the cluster Cepheid CF Cas, and compute a distance of 
$3160 \pm 160$ pc (distance modulus 12.5). Matthews et al. (1995) also 
calculate a mass of 5 M$_{\odot}$ for CF Cas. Adopting this as the 
main sequence turn-off (MSTO) mass then suggests an age of 60 - 70 Myr, although 
this is an upper limit given that the Cepheid progenitor may have experienced mass loss.

	The paper is structured as follows. The observations and the 
procedures used to produce the final images are described in \S 2, while the 
photometric measurements are discussed in \S 3. 
The CMDs and LFs of the clusters are presented in \S\S 4 and 5, while the 
spatial distributions of stars in the clusters are investigated in \S 6. 
A summary and discussion of the results follows in \S 7.

\section{OBSERVATIONS}

	The data were recorded with the CFHT MegaCam (Boulade et al. 2003) during the 
night of October 11, 2010. The MegaCam detector is a mosaic of 
thirty six $2048 \times 4612$ E2V CCDs that are arranged in 
a $4 \times 9$ format. A $\sim 1$ degree$^2$ field is imaged exposure$^{-1}$ 
with 0.185 arcsec pixel$^{-1}$ sampling. There are two 80 arcsec gaps between detector 
banks, and 13 arcsec gaps between individual CCDs.

	Images were recorded through $u*, g'$ and $r'$ filters. 
A single 60 sec exposure was recorded in $u*$, while single 
30 second exposures were recorded in $g'$ and $r'$. The angular resolution of 
the final images is 0.8 arcsec FWHM.

	The initial processing, which included bias subtraction and flat-fielding, was 
done with the CFHT ELIXIR pipeline. Each ELIXER-processed MegaCam image was parsed 
into six $3 \times 2$ CCD sub-mosaics for subsequent analysis. Experience has shown 
that such sub-mosaics are a manageable size for subsequent photometric analysis.

	Chromatic abberations in the MegaCam optics introduce filter-to-filter 
positional offsets that increase in size as one moves away from the center of 
the science field. Differential offsets on the order of an arcsec (i.e. a few pixels) 
were found between $u*$ and $g'$ images near the edge of the MegaCam field, while 
much smaller offsets were found between the edges of the $r'$ and $g'$ images. 
These distortions were removed by mapping the $u*$ and $r'$ images into the 
$g'$ image reference frame using the GEOMAP and GEOTRANS routines in IRAF.

\section{PHOTOMETRY}

\subsection{Photometric Measurements}

	Photometric measurements were made with the point spread function (PSF)--fitting 
routine ALLSTAR (Stetson \& Harris 1988). The input source catalogues, preliminary 
magnitudes, and PSFs were obtained by running tasks in the DAOPHOT (Stetson 1987) 
package. The PSFs were constructed from isolated bright stars, and 
faint companions were removed from these stars in an iterative manner using 
progressively improved versions of the PSF.

	ALLSTAR rejects sources that are not well-matched by the PSF and/or that 
hinder convergence. Still, sources that have poor photometric measurements 
and/or that are non-stellar may be present in the ALLSTAR output. 
As in previous studies (e.g. Davidge 2010), such objects 
were identified and removed using the magnitude error 
computed by DAOPHOT, $\epsilon$, as a guide.

	Stars that are neither saturated nor in crowded environments define a 
monotonic relation on the $\epsilon\ vs.$ magnitude diagram 
(e.g. Figure 2 of Davidge 2010), while resolved galaxies, unresolved blends, 
saturated stars, and cosmetic defects depart from this relation 
Following Davidge (2010), objects that did not fall on the unblended stellar sequence 
on the $\epsilon\ vs.$ magnitude diagram were deleted from the photometric catalogue. 
Objects with $\epsilon \geq 0.3$ magnitudes, most of which plumb 
the faint limits of the photometry, were also rejected.

\subsection{Calibration}

	Standard stars are observed during each MegaCam observing 
block. The zeropoints extracted from the standard star observations are then placed in 
image headers as part of ELIXER processing. This standard star information 
was used to transform the instrumental magnitudes into the SDSS $u'g'r'$ system. 

	The effective wavelengths of the $u*$ and $u'$ filters differ by $\sim 200\AA$, 
in the sense that $u*$ is at a longer wavelength, and this is a source of concern in the 
$u* \rightarrow u'$ transformation. Previous experience with $u*$ 
observations of the brightest main sequence stars in nearby 
galaxies found that a linear $u* \rightarrow u'$ transformation is adequate 
for these objects (e.g. Davidge \& Puzia 2011; Davidge et al. 2012). The veracity of 
the transformation process was checked in those studies by making comparisons with 
published photometry. Good agreement was also found between the predicted and 
observed locus of the upper main sequence on the $(g', u'-g')$ CMDs. The 
successful application of a linear $u* \rightarrow u'$ transformation in these 
studies is due to the modest size of the Balmer discontinuity in the spectral-energy 
distributions (SEDs) of massive upper main sequence stars. 

	For the present study, a linear $u* \rightarrow u'$ transformation was adopted.
Intrinsic colors were computed by assuming a single reddening for the 
entire field. Star-to-star reddening variations are to be expected 
because (1) the reddening material is not uniformly 
distributed across the field, and (2) the stars in the field are at a range of 
distances, and so are attenuated by different amounts of dust. 
However, experiments indicated that perturbing the adopted $E(B-V)$ by 
$\pm 0.05$ magnitudes did not affect greatly the transformed magnitudes.

	Clem et al. (2008) find that a linear $u* \rightarrow u'$ transformation may 
result in large uncertainties at intermediate spectral types, where the Balmer 
discontinuity is most pronounced, and so it is important to check the $u'$ 
magnitudes computed here. To do this, published $UBV$ photometry for King 12 and NGC 
7790 from Mohan \& Pandey (1984) and Alcala \& Arellano Ferro (1988) were transformed 
into $u'g'r'$ using the relations from Smith et al. (2002), and the results were 
compared with the MegaCam photometry. While CCD measurements are available for NGC 7790 
(e.g. Romeo et al. 1989), the photoelectric photometry of Alcala \& Arellano 
Ferro (1985) are prefered because they include the $U$ measurements that are essential 
for checking the transformed $u'$ magnitudes.

	In \S 4 it is shown that the MegaCam and transformed 
photoelectric measurements mesh together nicely on the 
$(g', u'-g')$ and $(r', g'-r')$ CMDs. Model isochrones also provide a reasonable 
match to the cluster sequences on the $(g', u'-g')$ CMDs over much of the main sequence 
(e.g. Figure 6). Still, recognizing the potential shortcomings in the 
$u* \rightarrow u'$ transformation, the $u'$ measurements play a subordinate role to 
the $g'$ and $r'$ data when computing cluster distance moduli and reddenings in \S 4.2.

\subsection{Data Characterization}

	Data completeness and the uncertainties in the photometry were assessed in 
two ways. First, artificial star experiments were run. The photometric properties 
and statistics of the recovered artificial stars can be used to measure the uncertainies 
in the photometry and calculate sample completeness in a direct way. Second, fields 
with densities that are comparable to those in the clusters were simulated by stacking 
sub-sections of the MegaCam images. A comparison of the photometric measurements made 
from the unstacked and stacked images can then be used to assess the impact of crowding 
in a differential manner.

	Artificial stars were assigned colors and brightnesses that fall along the 
field star main sequence. As with the actual observations, 
an artificial star was considered to be detected only if it was recovered in at 
least two filters (either $u' + g'$ and $g' + r'$). The results from ALLSTAR were 
filtered with the $\epsilon$-based criteria that were applied to the science frames.

	The artificial star experiments indicate that the data are complete when 
$g' \leq 22.5$ and $r' \leq 22$, while 50\% completeness occurs near 
$g' = 23.5$ and $r' = 23$. The random and systematic uncertainties in 
the photometry climb rapidly towards fainter magnitudes when sample 
completeness drops below 50\%, and so the point at which 50\% completeness occurs 
is a pragmatic estimate of the faint limit of data. This is demonstrated in 
the next section where it is shown that the 50\% completeness curve falls just above 
the lower envelope of data on the CMDs.

	The clusters have stellar densities that are a few 
times higher than in the field, and so the faint limit of the cluster photometry differs 
from that of the surrounding field. To assess the size of this difference, 
fields with stellar densities that are three times higher than is 
typical throughout most of the MegaCam image were simulated by stacking three $1000 
\times 1000$ pixel sub-sections of the MegaCam images. The image sub-sections 
selected for this experiment do not include the clusters. Source brightnesses were then 
measured in the stacked images using the procedures described in \S 3.1.

	The composite $(r', g'-r')$ CMD and $r'$ LF of objects in the three 
sub-fields are compared in Figure 1 with the CMD and LF obtained from the stacked images.
Significant differences between the two CMDs and LFs become evident between $r' = 20$ 
and 21. It is evident that tripling the stellar density affects the photometry when 
$r' > 20$, and so $r' = 20$ is adopted as the faint limit for the cluster observations.

\section{RESULTS: CMDs}

\subsection{A First Look at the CMDs}

	The $(g', u'-g')$ and $(r', g'-r')$ CMDs of King 12, NGC 7788, 
NGC 7790 are shown in the top panels of Figures 2 and 3. The 
King 12 and NGC 7788 CMDs include sources that are within 90 arcsec of the cluster 
centers, which is comparable to the cluster radii measured in \S 6.
The NGC 7790 CMD includes sources within 180 arcsec of the 
cluster center. While NGC 7790 extends out to larger radii 
(\S 6), it was decided to limit coverage to this region 
to reduce contamination from non-cluster members.

	The $\pm 2 \sigma$ uncertainties in $u'-g'$ and $g'-r'$ computed from the 
artificial star experiments are shown in the upper right hand panel of each figure. 
A line that corresponds to 50\% completeness is also shown in the NGC 7790 panels, 
and this falls just above the faint envelope of the NGC 7790 CMDs. 
The faint limit of these data is set by the $u'$ measurements in Figure 2, and 
the $g'$ measurements in Figure 3. While not shown in Figures 2 and 3, 
the artificial star experiments indicate that the data are 100\% complete 
roughly 1 magnitude above the 50\% completeness relation.

	Stars in the MegaCam images saturate near $g' \approx r' \approx 14$, and 
this produces a sharp cut-off at the bright end of the observations that prevents 
the MSTO from being sampled. Therefore, the cluster sequences in 
Figures 2 and 3 were extended to brighter magnitudes by adding photoelectric  
observations from Mohan \& Pandey (1984) for King 12 and Alcala \& Arellano 
Ferro (1988) for NGC 7790, which were transformed into the SDSS system as 
discussed in \S 3.2. Only stars with $g' < 14$ from these studies are 
shown, as there is a marked increase in scatter at fainter magnitudes among 
the photoelectric measurements from both studies. Corresponding observations of 
NGC 7788 made with linear detectors are not available in the literature.

	A prominent blue main sequence with $u'-g' < 1.5$ and 
$g'-r' < 1.0$ dominates the upper half of each cluster CMD. 
The vertical arm of the main sequence in the upper portions of the  
CMDs have similar $u'-g'$ and $g'-r'$ colors, which in turn suggests 
that the cluster reddenings are similar. This foreshadows one result of this 
paper -- that the reddening of NGC 7788 has heretofore been underestimated. 

	The clusters are at very low Galactic latitudes and 
-- with the exception of the brightest cluster members -- 
contamination from stars along the line of sight complicates efforts 
to identify cluster sequences on the CMDs. The extent of 
field star contamination is examined in the bottom rows of Figures 
2 and 3, where the CMDs of sources in annuli surrounding the clusters are 
shown. The control fields have an inner radius that is $1.5\times$ that of the 
radius used to construct the CMDs in the top rows of Figures 2 and 3, and an outer 
radius that allows the same total area on the sky as in the cluster to be sampled; 
thus, star counts in the CMDs in the lower panels of Figures 2 and 3 can be used to 
estimate directly field star contamination in the cluster CMDs. Given the large angular 
size of NGC 7790 (\S 6), some stars that belong to NGC 7790 are undoubtedly present 
in the control field. The relative numbers of blue main sequence 
stars in the CMDs of the various cluster and control fields suggest that the 
clusters have a stellar density that is $\sim 3 \times$ that in the surrounding field. 

	It is difficult to track the cluster sequences in the lower half of each 
CMD. This is because the cluster and field star sequences in the CMDs overlap at 
intermediate magnitudes. In addition, while the number of field main sequence 
stars grows towards fainter magnitudes as progressively larger volumes of the disk 
are sampled, the cluster LFs are such that the number of cluster stars per 
magnitude interval does not change greatly with $r'$ (\S 5). Finally, 
age plays a role in explaining the dearth of 
faint cluster members in the King 12 CMD, as the lower part of the 
main sequence is not yet populated in this young cluster (see below).

\subsection{Comparison with Main Sequence and Post-Main Sequence Models}

	The cluster CMDs are compared with isochrones from Girardi et al. (2004) 
in Figure 4. The distances and reddenings listed in Table 1 have been adopted 
for this initial comparison. The relations provided in Table 6 of Schlegel et 
al. (1998), which draw on the extinction laws of Cardelli et al. 
(1989) and O'Donnell (1994), were used to redden the models. Isochrones with Z = 0.019 
and ages of 10, 20, 40, and 80 Myr are shown. A 80 Myr sequence with Z = 0.030 
is also plotted to allow metallicity effects to be assessed. 

	A clear result from Figure 4 is that the reddening for NGC 7788 
in the WEDBA database is too small, as the models fail to match the vertical 
plume of main sequence stars with $r' < 15$. It is also apparent that the ages 
of these clusters are difficult to establish given the modest number of intrinsically 
bright stars. This holds even after including the photoelectric observations of 
bright stars in King 12 and NGC 7790. 

	The reddenings and distance moduli used in Figure 4 can be tuned to 
obtain better agreement between the models and observations, and the results of 
such an exercise are shown in Figure 5. An initial reddening 
for each cluster was selected so that the isochrones match the blue envelope of the 
vertical part of the main sequence with $r' < 15$ on the $(r', g'-r')$ CMD. 
The distance modulus was then adjusted with the reddening fixed and the best match 
with the lower envelope of the cluster sequence near the elbow in the cluster CMD 
near $r' = 16$ was found. A revised reddening estimate was then made, and the procedure 
was iterated until convergence.

	The revised distance modulus and color 
excess for each cluster are listed in Figure 5 above the cluster CMDs. 
The estimated uncertainties in the distance moduli are 
$\pm 0.1$, while the uncertainties in the reddenings are $\pm 0.05$ magnitude. 
The distance modulus found for NGC 7790 agrees with 
other recent estimates, which are summarized in Table 2. 
In contrast, the distance moduli of King 12 and 
NGC 7788 differ significantly from the values in the WEDBA database.

	While the $(g', u'-g')$ CMDs provide the tightest constraints on cluster 
ages (see below), preliminary ages can be obtained from Figure 5 for King 12 and 
NGC 7790. Based largely on the three stars with $r' < 11$ from Mohan \& Pandey (1984), 
the comparisons in Figure 5 suggest that King 12 has an age $< 40$ 
Myr. As for NGC 7790, the brightest star in the Alcala \& Arrelano Ferro (1988) 
observations is consistent with an age near 30 Myr, although the majority of fainter 
data points are consistent with an older age. An age can not be estimated for NGC 7788 
based solely on the MegaCam $g'$ and $r'$ data, although the situation is different 
when the $u'$ measurements are considered (see below).

	With the distance moduli and reddenings set from the $(r', g'-r')$ CMDs,
additional insights into the cluster ages can be obtained from the 
$(g', u'-g')$ CMDs, and isochrones are compared with the $(g', u'-g')$ CMD 
in Figure 6. The $u'$ photometry provides an 
independent check of the reddening, while the $u'-g'$ color is more 
sensitive to age and metallicity than $g'-r'$. This being said, the $u'-g'$ 
color can be difficult to reproduce in model isochrones, with uncertainties 
of up to a tenth of a magnitude (e.g. An et al. 2009). 

	The bluest stars in the $(g', u'-g')$ CMD of King 12 are consistent with an age 
$< 20$ Myr. For NGC 7790, the photoelectric measurements with $g' < 14$ suggest an 
age between 60 and 80 Myr. Finally, given the increased age sensitivity of $u'-g'$ 
among intermediate mass stars it is possible to estimate an age for 
NGC 7788, and the MegaCam data are consistent with an age of 20 -- 40 Myr. This 
is the least reliable age estimate in this study, and
photometry of brighter stars in this cluster are needed to confirm the 
age of NGC 7788.

\subsection{Comparisons with Pre-Main Sequence Models}

	Star clusters as young as King 12 and NGC 7788 should harbor a population 
of pre-main sequence (PMS) objects, and these will fall redward of the main sequence 
on the CMDs. In fact, some parts of the King 12 and NGC 7788 $(r', g'-r')$ CMDs, 
indicated in Figure 3, contain an excess number of objects when compared 
with the control fields. The area of the King 12 and NGC 7788 CMDs indicated 
in Figure 3 will be refered to as the `PMS box' in subsequent discussion.

	The PMS box in King 12 contains 40 objects, while there are 
32 objects in NGC 7788. For comparison, each of the control fields contain 
18 objects in the PMS box. The excellent agreement between 
the number density of these objects in the two control fields indicates that -- 
outside of the clusters -- the objects in the PMS box are uniformly distributed, as 
expected if they are non-cluster field objects. After subtracting the number of 
objects in the control fields, the content of the PMS box differs from zero 
at the $2.9-\sigma$ level in King 12, and the $2.0-\sigma$ level in NGC 7788. 

	The lower portions of the cluster CMDs are compared with models of 
PMS evolution from Tognelli et al. (2011) in Figure 7. 
There are considerable uncertainties in models of PMS evolution, and these have 
been reviewed by Siess (2001). The largest single source 
of uncertainty in the input physics is the recipe used to 
account for convection, although the equation of state, boundary conditions, 
rotation, and mass accretion will also affect the models. The translation of 
models onto the observational plane are also prone to uncertainties in the bolometric 
corrections, the T$_{eff}$ scale, abundances, and circumstellar extinction, although the 
latter likely affects the observational properties of PMS objects 
only for systems with ages $\leq 6$ Myr (Baume et al. 2003). 

	The luminosity $+$ effective temperature (T$_{eff}$) 
pairs tabulated by Tognelli et al. (2011) were transformed 
onto the observational plane using relations between T$_{eff}$, broad-band colors, and 
bolometric corrections derived from main sequence stars. The use of relations that 
are based on main sequence stars is not ideal, as the SEDs of PMS stars 
differ from those of main sequence stars. In addition, no allowance was made 
for circumstellar extinction, although the good agreement between 
age estimates based on PMS stars and more evolved main sequence stars in 
young cluster such as NGC 2362 (e.g. Moitinho et al. 2001) and NGC 3293 (Baume et al. 
2003) suggests that natal dust envelopes dissipate on timescales that are shorter 
than the ages of King 12 and NGC 7788. Indeed, observations and dynamical arguments 
both suggest that circumstellar disks likely dissipate over timescales of $\sim 6$ 
Myr (Mayne et al. 2007). 

	The properties of PMS models depend on factors including 
metallicity, mixing length ($\alpha$), helium abundance, and the 
initial deuterium abundance ($X_D$). The models in Figure 7 -- as well as those 
used later in this paper -- assume Z = 0.015, Y=0.278, 
$\alpha = 1.68$, and $X_D = 2 \times 10^{-5}$. Despite the inherent uncertainties in the 
transformation procedure -- coupled with the sensitivity to input physical parameters 
-- the PMS models pass through the portions of the King 12 and NGC 7788 CMDs that 
contain an excess population of red objects, suggesting that 
PMS objects have been detected in these clusters.

	The models also predict that PMS stars in 
NGC 7790 -- with an age $60 - 80$ Myr -- may also lurk in the 
MegaCam images. Such objects would be fainter (and hence more difficult 
to detect) than their closer and younger brethren in King 12 and NGC 7788. In fact, 
the NGC 7790 LF appears to be fully populated to $r' \sim 20$ (\S 5), suggesting that 
PMS stars in that cluster are fainter than $r' = 20$.

\section{RESULTS: LFs}

\subsection{Differential LFs}

	The $r'$ LFs of objects in King 12, NGC 7788, and NGC 7790 
are shown in Figure 8. The LFs are restricted to $r' < 20$, which 
is the magnitude range that the stacking experiments discussed in \S 3.3 indicate 
is not affected by crowding. Given this cut-off then main 
sequence stars in these clusters with masses $> 0.8 - 0.9_{\odot}$ are sampled.

	The LFs have been corrected for field star 
contamination by subtracting star counts over the entire MegaCam field, 
but excluding the areas occupied by the clusters. Drawing on 
number counts over such a large area reduces statistical errors, so that the 
random uncertainties in the final cluster LFs mainly reflect statistical 
flucuations in the cluster stellar contents. A potential shortcoming of 
estimating contamination from source counts that cover 
a large area near the disk plane is that reddening variations may affect the results.
Still, the overall trends of the background-subtracted 
LFs in Figure 8 do not change greatly if number counts taken from the control fields 
shown in Figure 3, which are less prone to large-scale reddening variations 
given their modest size and physical proximity to the clusters, 
are used instead of those based on the entire MegaCam field.

	The LFs of NGC 7788 and NGC 7790 in Figure 8 have broadly similar appearances, 
with a more-or-less flat section when $r' > 16$, with the break near $r' = 16$ due to 
an inflexion point in the mass-magnitude relation among main sequence stars. In 
contrast, there is an apparent deficiency of stars with $r'$ between 
16.5 and 19 in King 12 when compared with other clusters. The ratio of 
sources at the faint and bright end of the King 12 LF is also larger than in the 
other clusters. 

	The dotted line in Figure 8 is a model LF constructed from the Z = 0.019 
Marigo et al. (2008) sequences, which was generated using the interface at the 
{\it Padova database of evolutionary tracks and isochrones}
web site \footnote[2]{http://stev.oapd.inaf.it/cgi-bin/cmd}.
The model assumes an age of 30 Myr. While the clusters span a range 
of ages, it should be recalled that the MegaCam data sample 
only stars that are below the MSTO, and so age effects near the bright end of the LFs 
in Figure 8 should be negligible. A Kroupa (1998) IMF was adopted, and 
the model in each panel was scaled to match the cluster LF between $r' = 14.5$ and 16. 
The model LF does not include binaries. Simulations indicate 
that the feature used below to estimate cluster ages -- the lower 
main sequence cut-off -- is not sensitive to binarity (Cignoni et al. 2010).

	The model LF does not match the King 12 observations very well, 
with number counts at some intermediate magnitudes differing from the model at more 
than the $2\sigma$ level. Such a disgreement is perhaps not unexpected, as the main 
sequence will not be fully populated at intermediate masses in young clusters like King 
12, and the magnitude at which such departures occur provides information about 
cluster age. Mayne et al. (2007) compare main sequence and PMS sequences in a number of 
open clusters, and identify a gap that is associated with the transition from a fully 
convective PMS object to one with a radiative core. The 
rate at which objects traverse this gap depends on 
the mass of the protostar, in the sense that more massive stars traverse the gap faster.
The gap will appear as a discontinuity in the cluster LF.

	The model provides a better match to the LFs 
of NGC 7788 and NGC 7790, and the overall agreement between 
the model and the cluster LFs near the bright end provides a (loose) 
consistency check on their distance moduli. Still, the NGC 
7788 LF departs from the model LF at $r' \sim 19$, or M$_{r'} \sim 5.5$. The 
simulations discussed in \S 3.3 suggest that this is not due to incompleteness. 

\subsection{Integrated LFs}

	The integrated LFs of the clusters will provide additional 
insights into any differences between the observed and model brightness distributions 
in King 12 and NGC 7788. Therefore, the differential LFs shown in Figure 
8 were integrated, and the differences between the observed and model LFs 
of these clusters -- $\Delta_{ilf}$  -- are shown in Figure 9. 
A deficiency in the observed number of stars with respect to the model 
produces a positive value of $\Delta_{ilf}$.

	In the case of King 12, $\Delta_{ilf}$ differs from zero with a significance 
in excess of the $2\sigma$ level at $r' = 16.5$, and at fainter $r'\ \Delta_{ilf}$ is 
persistently positive. This result, coupled with the comparison between the 
diffential LFs of King 12 and the model in Figure 8, suggests that the onset of the PMS 
occurs at $r' = 16.25 \pm 0.5$ in King 12. This corresponds to M$_{V} \sim 3 
\pm 0.25$, and the mass-magnitude relation for main sequence stars shown in Table VIII of 
Habets \& Heintze (1981) indicates that main sequence 
stars with this M$_V$ have a mass $1.4 \pm 0.1$ M$_{\odot}$. 

	This lower mass limit for main sequence stars can be used to estimate the age 
of King 12. Adopting the same parameters as those used to generate the PMS isochrones 
shown in Figure 7, the Tognelli et al. (2011) models predict that a star with a 
mass $1.4 \pm 0.1$ M$_{\odot}$ will take $\sim 15 \pm 2$ Myr to relax onto the main 
sequence, and this is adopted as the PMS-based age for King 12. The 
age calculated from the main sequence cut-off thus falls within the 
range that was obtained from the CMDs in \S 4.2. 
With such an age, the red objects at the faint end of 
the King 12 LF in Figure 8 are likely to be PMS stars, and in \S 6 it is 
demonstrated that the spatial distribution of 
these objects is consistent with cluster membership. 

	There is a substantial increase in $\Delta_{ilf}$ near 
$r' = 17.25$ in the upper panel of Figure 9 -- could this 
be the onset of the PMS in King 12 rather than the more modest feature 
discussed above? In order for this to be the case then the 
non-zero $\Delta_{ilf}$ at $r' = 16.25$ would have 
to be a statistical fluke. If the onset of the PMS occurs at $r' = 17.25$ then 
stars as massive as 1.2 M$_{\odot}$ are on the King 12 main sequence. The 
age of King 12 based on the main sequence cut-off is then $22 \pm 2$ Myr, which 
still agrees with the upper end of the range predicted by bright main sequence stars.

	The evidence for the onset of a PMS sequence in the LF of NGC 7788 is 
not as robust as in King 12. It is evident from the bottom panel of Figure 9 that the 
the observed and model integrated LFs of this cluster 
may start to diverge near $r' = 19$, although 
$\Delta_{ilf}$ only differs from zero at the $1.5-\sigma$ level 
at this $r'$. Still, it is encouraging that $\Delta_{ilf} > 0$ 
when $r' \leq 19$, and so $r' = 18.75 \pm 0.25$ is adopted as the tentative onset of the 
PMS in NGC 7788. This corresponds to M$_V \sim 5.6 \pm 0.25$, and 
the Habets \& Heintze (1981) calibration indicates that a main sequence star with 
this M$_V$ has a mass $0.86 \pm 0.04$ M$_{\odot}$. The Tognelli et al. (2011) models 
predict that such an object will take $\sim 47 \pm 3$ Myr to relax onto the main 
sequence. This age estimate favors the old end of the age range for NGC 7788 
measured in \S 4.2.

	There is not a statistically significant dfference 
between the observed and model LFs of NGC 7790 when $r' \leq 20$, 
which corresponds to M$_{r'} \leq 5.5$. This suggests that the main sequence of this 
cluster is populated by stars with masses $\geq 0.9$ M$_{\odot}$. The Tognelli et al. 
(2011) models then predict that NGC 7790 should be older than $\sim 30$ Myr, and 
this is consistent with the age of this cluster estimated in \S 4.2.

\section{CORRELATION FUNCTIONS AND CLUSTER STRUCTURE}

	The large-scale clustering properties of objects in King 12, NGC 7788, 
and NGC 7790 are investigated using the two-point correlation function (TPCF). The TPCF 
is based on a simple observable - the separation between 
two points on the sky -- and so the interpretation of the TPCF is not complex. 
Most previous applications of the TPCF to star clusters examined 
spatial scales that sample the transition between multiple star systems 
(binaries, triples, etc) and the larger-scale clustering of 
such systems (e.g. Gomez et al. 1993; Larson 1995, Simon 1997). 
Larson (1995) finds that the clustering properties of stars in the Taurus and 
Ophiucus star-forming regions changes at separations $\sim 0.04$ parsecs, and 
notes that this spatial scale is comparable to the Jeans length in molecular clouds. 
The separation at which the correlation properties change can then be interpreted as 
the maximum spatial extent of individual stellar systems that form within a 
single proto-stellar envelope. Power in the correlation spectrum at larger 
spatial scales is due to clustering between individual star systems.

	Three samples of objects are defined for this investigation. 
The first consists of bright main sequence stars (BMSs), which have $r'$ between 
14 and 16, and $g'-r'$ between 0 and 0.7. The second 
consists of red plume sources (RPSs), which have $r' < 19$ and 
$g'-r'$ between 1.2 and 2.5. The extraction region for the RPS sample was 
selected to avoid possible PMS stars in King 12 and NGC 7788 (\S 5). The third sample 
consists of candidate PMS objects. This sample is restricted to King 12 and NGC 7788, 
and contains objects in the PMS region of the $(r', g'-r')$ CMD indicated in Figure 3.

	These samples probe very different populations. The 
intermediate mass main sequence stars that make up the BMS sample 
are present in all three clusters (e.g. Figure 3), and so provide a 
homogeneous basis for comparing cluster properties. The objects in the PMS sample 
are a mix of cluster and non-cluster objects, and this complicates efforts 
to compare their spatial distribution with other objects (see below). As for the RPSs, 
these isotropically distributed objects are not cluster 
members, and so are used to normalize the TPCFs. 
To the extent that the sources in the BMS, PMS, and RPS samples occupy 
comparable areas of the science detector, then this normalization corrects 
for issues arising from detector geometry, such as gaps between detectors 
and finite area coverage.

\subsection{Main Sequence Stars}

	The TPCFs obtained by ratioing the separation functions (SFs) of BMS and RPS 
sources within a 9 arcmin radius of the three clusters are shown in Figure 10. The 
curves in Figure 10 have been normalized based on the number of objects in each sample. 
The edge of the MegaCam science field is 6 arcmin from the center of NGC 7790, 
and so there is incomplete spatial coverage of this cluster for separations 
$\geq 12$ arcmin. However, as both the BMS and RPS samples are vignetted, 
then the impact of incomplete spatial coverage on the shape of the TPCF is removed. 
While there is not a discontinuity near 12 arcmin separations in the NGC 7790 TPCF, 
the shape of the TPCF of NGC 7790 at separations in excess of 12 arcmin will
be affected by incomplete spatial coverage if the cluster is not 
azimuthally symmetric on the sky - that is, any sub-structures in the southern 
part of the cluster will be missed. 

	Following Simon (1997), jackknife re-sampling was used to assess uncertainties 
in the correlation functions. This approach automatically tracks the influence of 
stars over the full range of separations. Results of re-sampling experiments for NGC 
7788 are shown in the lower panel of Figure 10. Each realization shows the TPCF 
that is obtained after randomly extracting 25\% of the sources in the BMS 
and RPS samples. 

	A number of broad conclusions can be reached from the re-sampling experiments.
First, the dispersion amongst the TPCFs increases towards 
smaller separations. This is to be expected given that the number of possible 
object-object pairings drops as one moves to smaller separations, leading to 
greater sensitivity to stochastic flucuations. A similar affect occurs at very 
large separations. Second, the dispersion in the re-sampled NGC 7788 TPCFs is comparable 
to the bin-to-bin chatter in the full NGC 7788 sample. The local bin-to-bin 
chatter thus provides an estimate of the random uncertainty in the TPCF at a given 
separation. Finally, the overall shapes of the TPCFs in the lower panel of Figure 10
are similar. This indicates that the structural features identified 
in the TPCF of NGC 7788 (see below) are robust detections. 

	If the separation distributions of two 
uniformly distributed populations are ratioed then the result will be a TPCF 
that is flat, with bin-to-bin noise introduced by stochastic variations 
in the spatial distributions of the stars. The TPCFs of the clusters in Figure 10 
are not flat, indicating that the BMS objects are not uniformly distributed. Rather, 
the TPCFs of all three clusters show a general trend of increasing coherence towards 
smaller separations, which is a consequence of the central concentration of stars in 
the clusters. However, this broad similarity aside, the TPCFs of King 12 and NGC 
7788 differ from those of NGC 7790, and this is likely due to structural differences. 

	The TPCF of an isolated cluster with a well-defined boundary 
that is surrounded by a uniformly distributed field population will -- at separations 
that exceed the cluster diameter -- be flat out to the largest separations probed by 
the sample. This is clearly not the case for King 12 and NGC 7788, as the 
TPCFs of these clusters outside of their central peaks have 
distinct plateaus. The first plateau, labelled `Halo' in the NGC 7788 
panel, likely results from a population of BMS objects that are not as tightly 
clustered as objects in the main body of the star cluster, but still have 
a source density that exceeds that at larger separations. The halo component 
extends out to 10 arcmin separations (5 arcmin radius) around both clusters. 
The TPCF then drops to a second plateau at larger separations, labelled `Field' in 
the NGC 7788 panel, where the BMS and RPS objects are again uniformly distributed, 
but with a lower density than in the halo. The inflexion points at small separations, 
where the peak in the cluster TPCF blends into the surrounding halo, indicate that the 
main body of King 12 has a radius of $\sim 1$ arcmin, 
while the main body of NGC 7788 has a radius of 1.5 arcmin. 

	The TPCF of NGC 7790 does not have 
the prominent peak at small separations that is the signature of a central 
density concentration; rather, the TPCF of NGC 7790 is flat 
at separations $< 5$ arcmin. Given that the separation statistics are dominated by 
cluster objects, then this suggests that BMS objects are  
more-or-less uniformly distributed within 2.5 arcmin of the 
center of NGC 7790. The TPCF of NGC 7790 drops steadily for separations 
between 5 and 15 arcmin, and a uniform distribution of BMS stars is encountered 
at separations in excess of 15 arcmin. For comparison, 
Gupta et al. (2000) measure a core radius of 3.7 arcmin for 
NGC 7790, and find that cluster stars largely disappear at $1.5\times$ this value. 

	There is no evidence of a halo around NGC 7790. If NGC 7790 has a 
uniform-density halo then it must be located at much larger distances from the cluster 
center than is the case for the other two clusters. In addition, the NGC 7790 TPCF at 
large radii has a level of coherence that is comparable to the TPCFs of the field regions 
around the two other clusters. If there is a halo surrounding NGC 7790 then it 
must have a much lower density relative to the main body of the cluster than in 
either King 12 or NGC 7788.

\subsection{PMS Objects}

	The clustering properties of objects in the King 12 and NGC 7788 PMS samples are 
investigated in Figure 11. The top row of this figure shows the TPCFs obtained by 
ratioing the SFs of the PMS and RPS samples, while the lower row shows the 
ratio of the SFs of PMS objects and BMS stars in each cluster. 
With the possible exception of separations $< 1$ arcmin, PMS objects 
in King 12 have a roughly uniform distribution for separations $< 9 - 10$ arcmin. At 
larger separations the degree of clustering drops. 
This is similar to the clustering properties of BMS stars in King 12, 
providing additional evidence that the red sources detected in King 12 are PMS sources. 

	PMS stars should have a lower mass than objects 
that are already on the main sequence. That PMS stars 
are clustered out to 10 arcmin separations in King 12 indicates that the halo around this 
cluster (1) is populated by stars having a range of masses, and (2) has 
a radial extent that is similar for objects of all masses. Still, the ratio of the 
PMS and BMS SFs in the lower left hand panel of Figure 11 
indicates that the objects in these samples have different clustering properties. This 
may be due in part to different levels of contamination from 
non-cluster sources in each sample.

	In contrast to King 12, the TPCF of PMS objects in NGC 7788 is 
more-or-less flat over all separations, indicating that the candidate PMS objects 
are uniformly distributed, with no evidence of clustering. The ratio of PMS 
and BMS SFs in the vicinity of NGC 7788, shown in the lower right hand panel 
of Figure 11, indicates that the objects in these samples have very different 
clustering properties. The size of the PMS sample in NGC 7788 has a lower statistical 
significance than in King 12 ($2\sigma$ versus $3\sigma$ -- \S 
4.3), and there is a higher fractional contribution from non-cluster objects. 
Contamination from non-cluster objects, which are uniformly distributed,  
will cause the TPCF to become flatter. A deep emission line survey of NGC 7788 will help 
assess the spatial distribution of any PMS stars in this cluster.

\section{SUMMARY AND DISCUSSION}

	Images obtained with the CFHT MegaCam have been used to investigate 
the young star clusters King 12, NGC 7788, and NGC 7790. 
These clusters are at distances of 2 -- 3 kpc and have angular sizes of 
at least a few arcmin. They are also close together on the sky, and all three 
can be observed with a single MegaCam pointing. 
Neither King 12 nor NGC 7788 have been extensively studied in the past.

	The properties of main sequence stars and supected proto-stars have been 
explored in these clusters. In particular, the MegaCam data have been used (1) to 
estimates ages based on stars that are on the upper and lower parts of the main sequence, 
(2) to estimate distances, and (3) to explore the spatial distribution 
of stars in these clusters. The properties of faint main sequence and pre-main sequence 
stars, the implications of the cluster distances and ages on a possible common 
origin, and the structural properties of the 
clusters are discussed in turn in the following sections.
 
\subsection{Main Sequence and Pre-Main Sequence Stars}

	Ages have been estimated for all three clusters using the photometric properties 
of stars near the MSTO. Comparisons with isochrones suggest that NGC 7790 has an 
age 60 -- 80 Myr, and this is consistent with the age infered from the mass of CF Cas 
(Matthews et al. 1995), with the assumption that the Cepheid has experienced 
no significant mass loss during its evolution (\S 1). The properties 
of stars near the MSTOs of King 12 and NGC 7788 suggest an age $< 20$ Myr for the 
former and 20 -- 40 Myr for the latter.

	The MSTO is not an ironclad chronometer for low mass clusters, as the morphology 
of cluster sequences near the MSTO on the CMDs of such systems are susceptible to 
stochastic effects. Unresolved binaries will also complicate efforts to identify the 
MSTO. Stochastics effects are much less of an issue for lower mass cluster members, 
as these objects are more numerous than the most massive cluster members. For this 
reason, the main sequence cut-off is a potentially powerful alternate means of measuring 
ages in young, spatially diffuse, low mass clusters. 

	The use of the main sequence cut-off to estimate cluster ages is 
not new, and Belikov et al. (1997), Mayne et al. (2007), 
and Cignoni et al. (2010) have used discontinuities in the magnitude distribution 
of cluster stars to estimate ages. The amplitude of the discontinuities in 
LFs depends on cluster age (e.g. Mayne et al. 2007). The feature found by Belikov et al. 
(1997) in the LF of the Pleiades, which has an age 
$\sim 100$ Myr, occurs near M$_V = 6$ and has a modest size, whereas 
the models examined by Cignoni et al. (2010) predict 
features of a much more substantial size for systems with ages 
$\sim 10$ Myr. Cignoni et al. (2010) suggest that 
main sequence cut-off/PMS onset signatures in LFs are most 
useful as age estimators in systems that are younger than $< 30$ Myr. 

	The main sequence cut-off is not without its difficulties as an age indicator. 
There are uncertainties in the models of PMS evolution (e.g. Siess 2001), and 
the identification of the lower end of the main sequence may be complicated by 
contamination by non-cluster members. Another potential complication is that the 
stellar content of a cluster may not be coeval, and age dispersions of a few Myr may be 
present (e.g. Da Rio et al. 2010). The PMS stars that are only now relaxing onto 
the main sequence may then be a few Myr younger than the more massive stars near 
the MSTO. This being said, good agreement between upper 
and lower main sequence age estimates have been found in 
studies of NGC 2362 (Moitinho et al. 2001; Dahm 2005; Delgado et al. 2006) 
and NGC 3293 (Baume et al. 2003).

	Ages have been estimated for King 12 and NGC 7788 using the 
properties of the faintest main sequence stars. 
Given the various uncertainties in models of PMS evolution (\S 4.3), it is 
encouraging that the ages estimated for these cluster from features 
at intermediate and faint magnitudes in the $r'$ LF do not differ greatly from those 
obtained from stars near the MSTO. For King 12, the 
main sequence cut-off near $r' = 16.25$ yields an age $\sim 15 \pm 2$ Myr, which 
is consistent with the MSTO age estimate of $< 20$ Myr. Even 
if the main sequence cut-off in King 12 is 1 magnitude fainter (\S 5), then 
the age estimated for King 12 increases by only 7 Myr. As for NGC 7788, 
the lower end of the main sequence yields an age $\sim 47 \pm 3$ Myr, 
whereas upper main sequence stars give an age in the range 20 -- 40 Myr. 

	We close this section by noting that candidate PMS stars 
have been detected in both King 12 and NGC 7788. The detection of PMS 
stars is most secure in King 12, as these objects 
have large-scale distributions that are similar 
to those of bright main sequence stars (\S 6). If these objects 
are PMS stars then they should be sources of H$\alpha$ emission, and this can be 
checked directly by obtaining moderately deep narrow-band images of each cluster. 
A population of PMS stars should also be present in NGC 7790, although these 
objects likely have $r' > 20$, and so are below the faint limit of the MegaCam data.

\subsection{Cluster Distances and Kinship}

	Prior to the current study, some properties of King 12, NGC 7788, and NGC 7790 
hinted at a possible common origin, either in the same GMC or star-forming complex. 
In particular, the projected separations of these clusters is only $\sim 30$ parsecs, and 
previous studies yielded ages that agree to within a few tens of Myr. Neither 
the projected distribution of the clusters on the sky nor the large age range 
rule out an origin in the same GMC. Random motions in molecular 
clouds are a few km sec$^{-1}$ (Solomon et al. 1987), and so clusters that formed in 
the same GMC might disperse over $\sim 100$ parsecs after 30 Myr. 
In addition, a typical GMC may survive for $27 \pm 12$ Myr (roughly three 
free-fall times -- Murray 2011), suggesting that GMCs have the 
potential to form clusters over timescales of tens of Myr. 

	The possibility that these clusters may have 
originated in the same GMC -- and the prospect of probing the SFH of that GMC -- 
was the prime motivation for obtaining the MegaCam images used in this study. 
However, the distances measured in the current work rule out 
these clusters forming in the same GMC. King 12 and NGC 
7788 are separated by $\sim 300$ parsecs along the line of sight 
whereas NGC 7790 is $\sim 0.7$ kpc more distant than King 12. Still, while an origin 
in the same GMC is ruled out, these clusters may have formed in the same star-forming 
complex. Indeed, large star-forming complexes may extend over many hundred of parsecs 
(e.g. review by Efremov 1995), and can have SFHs that 
span many tens of Myr (e.g. Garcia-Benito et al. 2011).

\subsection{Cluster Structure}

	While the TPCF has not traditionally been used to probe the {\it large-scale} 
structure of star clusters (but see Sanchez \& Alfaro 2009), 
it is used here to investigate the distribution of 
stars in all three clusters out to separations of 20 arcmin. 
The TPCF multiplexes information from all objects in a sample, 
thereby making efficient use of the available information. The TPCF is also based on a 
simple observable -- the separation between two points on the sky -- providing an 
intuitive basis for interpreting the results. The star-star 
separation function (S3F) is a related statistic, which 
Davidge et al. (2011) used to investigate the diffusion 
timescales of stars from star-forming complexes in the outer disk of M33.

	The TPCFs of King 12 and NGC 7788 indicate that BMS stars in both clusters 
have compact cores that are confined to the central $\sim 0.1$ parsecs 
of each cluster. The main bodies of these clusters are in turn surrounded by  
BMS stars that have a uniform distribution on the sky. These halos 
extend out to $\sim 5$ arcmin from the cluster centers, which is a few 
times the radius of the main body of the clusters.

	As the oldest cluster, NGC 7790 is likely to be more dynamically evolved than 
King 12 and NGC 7788, which in turn may affect the core regions. 
Bastian et al. (2008) investigate the structural properties of young star 
clusters, and find that the structure of the core may change with time, with 
clusters having ages $< 20$ Myr showing the greatest range of structural diversity when 
compared with younger clusters. Indeed, Sanchez \& Alfaro (2009) find that clusters 
that are as old as 100 Myr may still contain sub-structures 
that have not yet been obliterated by dynamical processes, although other clusters 
with this age show no evidence for such sub-structures. 
The TPCF of NGC 7790 clearly differs from those of King 12 and NGC 7788, in that 
NGC 7790 has a much larger angular size and there is no obvious central core. If 
NGC 7790 once had a core then it has since dissipated.

	Where did the BMS stars in the halos that surround King 12 and NGC 7788 
originate? The extended halos are probably not the product of 
dynamical relaxation. Signatures of mass segregation might be expected 
to develop after a few cluster crossing times, and Prisinzano et al. (2003) find evidence 
for mass segregation in the $\sim 100$ Myr cluster NGC 2422, in the form of a spatially 
diffuse population of low mass main sequence stars.
The LFs of main sequence stars in NGC 7788 and NGC 7790 are similar in shape over 
most of the magnitude range probed here, and are well matched 
by models that are based on the Kroupa (1998) solar neighborhood mass function. 
That the LFs of NGC 7788 and NGC 7790, which have different ages and physical 
sizes, are consistent with the solar neighborhood mass function suggests 
that dynamical evolution has not yet affected the spatial distribution of objects 
with masses $> 0.6 - 0.7$ M$_{\odot}$ in these clusters. 
This indicates that the BMS stars that are used to 
construct the TPCF, which are among the most massive stars in these clusters, 
have not been displaced to large radii by mass segregation.

	The possibility that the halo stars may have formed {\it in situ} -- in the 
peripheral regions of the density enhancement that evolved into the main bodies of 
the clusters -- seems unlikely. With their uniform density distribution, 
the halos are structurally distinct from the central 
cluster. Star formation has a fractal distribution over spatial scales 
in excess of 10 parsecs (e.g. Gomez et al. 1993; references in Elmegreen \& 
Elmegreen 2001), and there is no evidence for the formation of distinct, large-scale 
halo components in star-forming environments. 

	We suggest that the halo stars aroung King 12 and NGC 7788 originated 
at smaller radii in the clusters and migrated outwards, possibly in response to the rapid 
expulsion of large quantities of star-forming material earlier in the evolution of the 
cluster. While once thought to be catastrophic for cluster survival, simulations 
suggest that the loss of substantial amounts of gas may 
not lead to the immediate disruption of a young cluster. Rather, stars that disperse 
from a cluster due to such an event may linger for $10+$ Myr in a low surface brightness 
region surrounding the cluster remnant (e.g. Bastian \& Goodwin 2006).
With an age in excess of 30 Myr then if NGC 7790 once had a halo of this nature 
then it may have by now dissipated into the surrounding field (Bastian \& Goodwin 2006). 

	The large-scale spatial distributions of objects that have very different masses 
may provide clues into the dynamical state of the cluster at the time of gas 
loss. Objects in the King 12 PMS and BMS samples have similar 
clustering properties, with a uniform distribution of objects in both samples 
for separations between 5 and 10 arcmin (Figure 11), and a drop in coherence 
for separations $> 10$ arcmin. Despite having very different masses, 
the similarity in clustering properties at separations 
associated with the cluster halo suggests that the objects in the PMS and BMS samples 
had similar kinematic properties at the time of gas ejection. This is consistent 
with King 12 still experiencing dynamical evolution at the time of gas loss, in agreement 
with simulations of early cluster evolution discussed by Delgado et al. (2011). 

\acknowledgements{It is a pleasure to thank the anonymous referee for comments that 
greatly improved the paper.}

\clearpage

\begin{table*}
\begin{center}
\begin{tabular}{cccc}
\tableline\tableline
Name & E($B-V$) & $\mu_0$ & log(t) \\
\tableline
NGC 7788 & 0.28 & 11.91 & 7.6 \\
NGC 7790 & 0.55 & 12.48 & 7.7 \\
King 12 & 0.60 & 11.98 & 7.0 \\
\tableline
\end{tabular}
\end{center}
\caption{Cluster Properties from the WEDBA Database (August 2012)}
\end{table*}

\clearpage

\begin{table*}
\begin{center}
\begin{tabular}{ccc}
\tableline\tableline
E($B-V$) & $\mu_0$ & Reference \\
\tableline
$0.51 \pm 0.03$ & $12.6 \pm 0.15$ & Gupta et al. (2000) \\
$0.54 \pm 0.05$ & $12.45 \pm 0.10$ & Lee \& Lee (1999) \\
--  & $12.48 \pm 0.11$ & Mathews et al. (1995) \\
$0.54 \pm 0.04$ & $12.65 \pm 0.15$ & Romeo et al. (1989) \\
& & \\
$0.56 \pm 0.05$ & $12.51 \pm 0.10$ & Current Study \\
\tableline
\end{tabular}
\end{center}
\caption{Recent Color Excesses and Distance Moduli of NGC 7790}
\end{table*}

\clearpage

\begin{figure}
\figurenum{1}
\epsscale{1.00}
\plotone{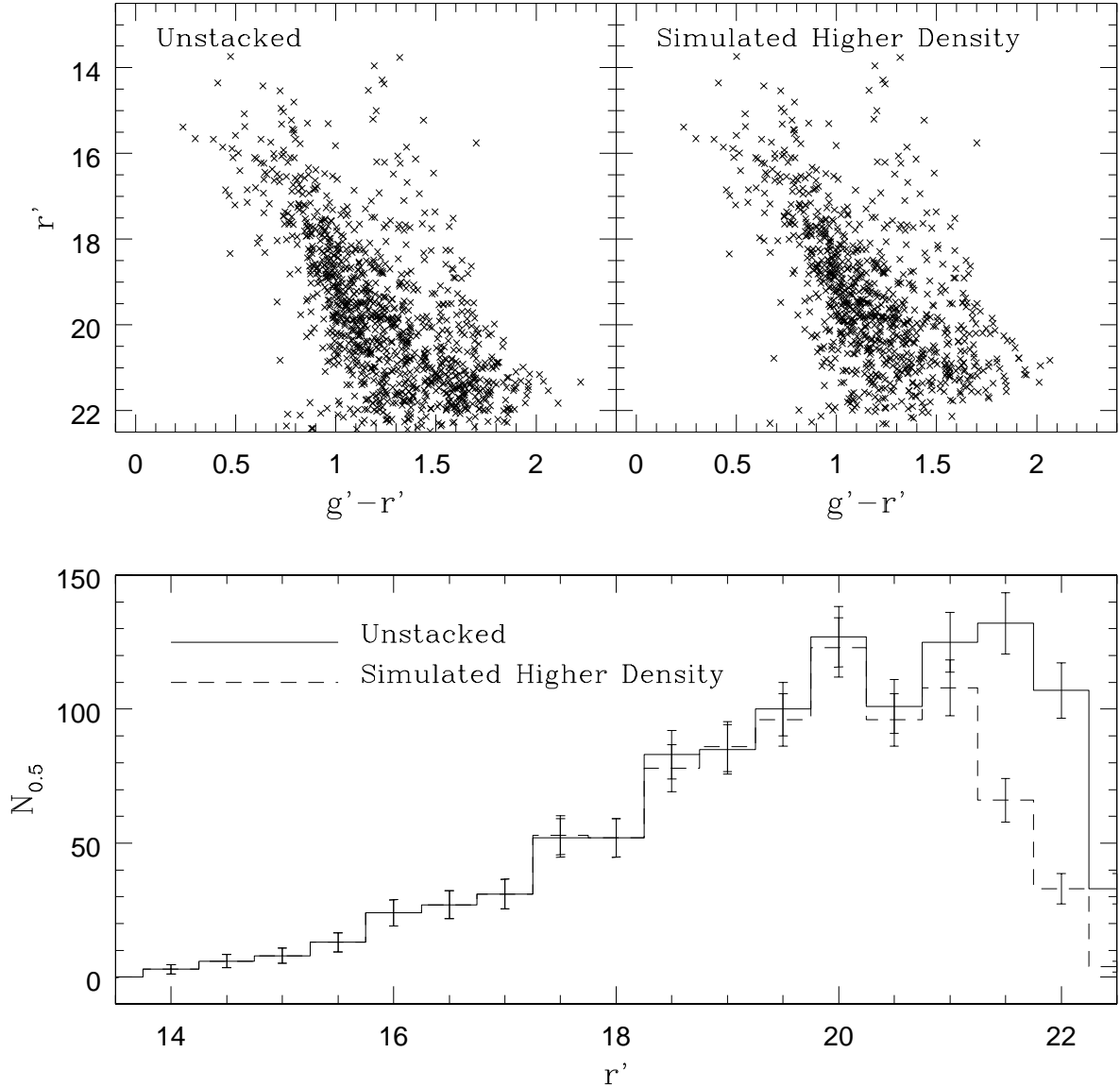}
\caption{The results of stacking experiments that were run to examine 
the affect of stellar density on the photometric faint 
limit. The composite $(r', g'-r')$ CMD of stars in three $1000 
\times 1000$ pixel areas that avoid the star clusters is 
shown in the left hand panel of the top row. The CMD in the right hand panel 
was measured from the sum of the images of the same three areas, 
thereby simulating a field with a density that is $3 \times$ higher than in the 
field, which is comparable to the stellar density in the clusters. 
The bottom panel compares the composite $r'$ LF of the three areas (solid 
line) with the LF measured from the stacked images (dashed line). The error bars 
show uncertainties due to counting statistics. Significant differences between the 
CMDs and LFs obtained from the stacked and unstacked images occur between  $r' = 20$ and 
21, and so $r' = 20$ is adopted as the faint limit for the cluster photometry.}
\end{figure}

\clearpage

\begin{figure}
\figurenum{2}
\epsscale{0.75}
\plotone{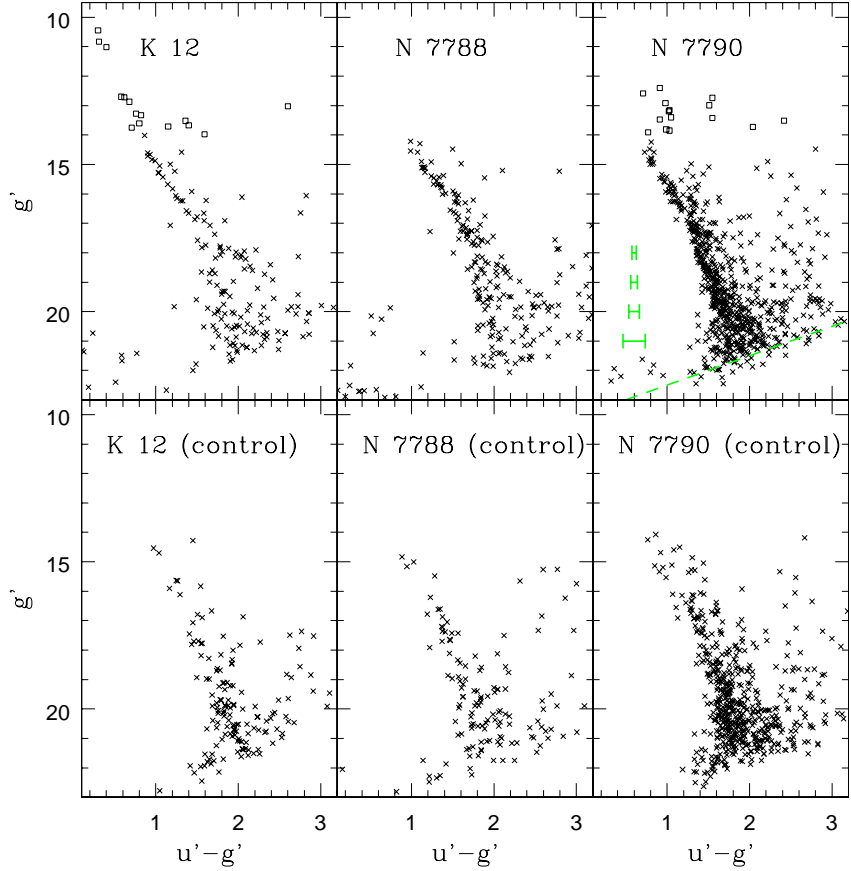}
\caption{The $(g', u'-g')$ CMDs of stars in King 12, NGC 7788, and NGC 7790. The CMDs 
in the top panel sample regions within 90 arcsec of the centers of King 12 and NGC 7788, 
and within 180 arcsec of the center of NGC 7790. The open squares are measurements 
with $g' < 14$ from Mohan \& Pandey (1984) and Alcala \& Arellano Ferro (1988) that 
have been transformed into the SDSS system. The CMDs of sources in an annulus surrounding 
each cluster that samples the same total area on the sky as the cluster CMD -- and so 
provides a rough estimate of field star contamination -- are shown in the bottom row. 
The $\pm 2 \sigma$ uncertainties in $u'-g'$ estimated from the artificial star 
experiments are shown in the NGC 7790 panel at various $g'$ magnitudes. The 
$2\sigma$ dispersion predicted by the artificial star experiments 
more-or-less matches the observed width of the NGC 7790 main 
sequence. The dashed green line is the 50\% completeness limit predicted 
by the artificial star experiments, which is defined by completeness in the 
$u'$ measurements (i.e. the $g'$ measurements are complete over the plotted 
magnitude range). 100\% completeness occurs roughly 1 magnitude above the 
50\% relation. With the caveat that modest numbers of cluster stars may be present in 
these control fields, it is evident that field star contamination is significant, 
especially in the lower half of the cluster CMDs.}
\end{figure}

\clearpage

\begin{figure}
\figurenum{3}
\epsscale{0.75}
\plotone{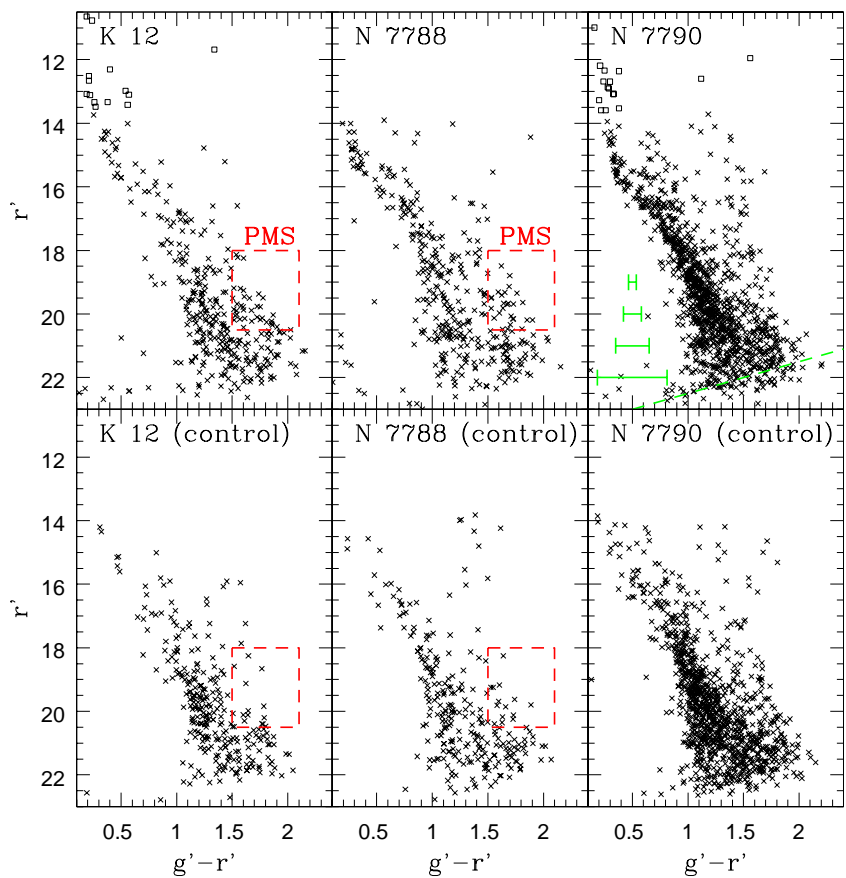}
\caption{The same as Figure 2, but showing the $(r', g'-r')$ CMDs of the clusters 
and surrounding fields. The red boxes indicate regions of the King 12 and NGC 
7788 CMDs that contain a population of objects that is not seen in the 
field. It is argued that the majority of these sources are PMS objects.}
\end{figure}

\clearpage

\begin{figure}
\figurenum{4}
\epsscale{1.00}
\plotone{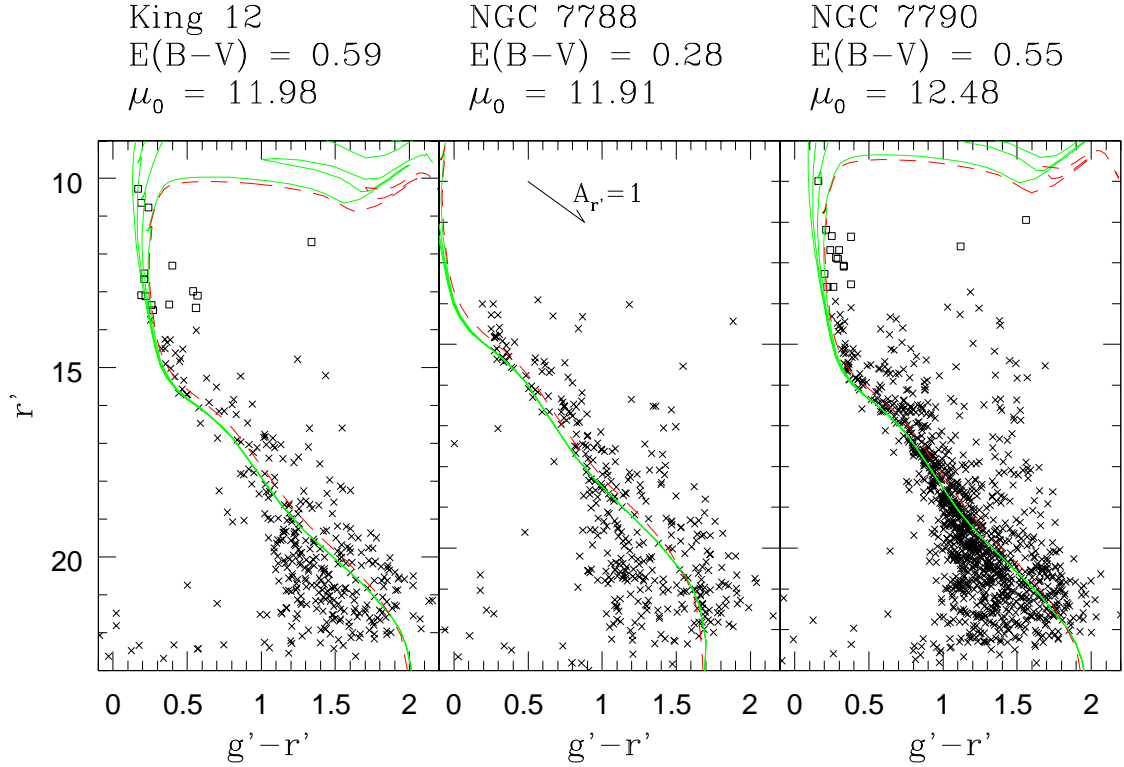}
\caption{The $(r', g'-r')$ CMDs of King 12, NGC 7788, and NGC 7790 are compared with 
isochrones from Girardi et al. (2004). The distances and reddenings listed 
in Table 1 have been used. The solid green lines are Z=0.019 sequences 
with ages of 10, 20, 40, and 80 Myr; a Z = 0.030 sequence with an 
age of 80 Myr is shown as a dashed red line. A reddening vector with a length 
that corresponds to A$_{r'} = 1$ mag is shown in the NGC 7788 panel. A 
higher reddening than that listed in the WEDBA database is required for the 
models to match the color of main sequence stars with $r' < 15$ in NGC 7788. }
\end{figure}

\clearpage

\begin{figure}
\figurenum{5}
\epsscale{1.00}
\plotone{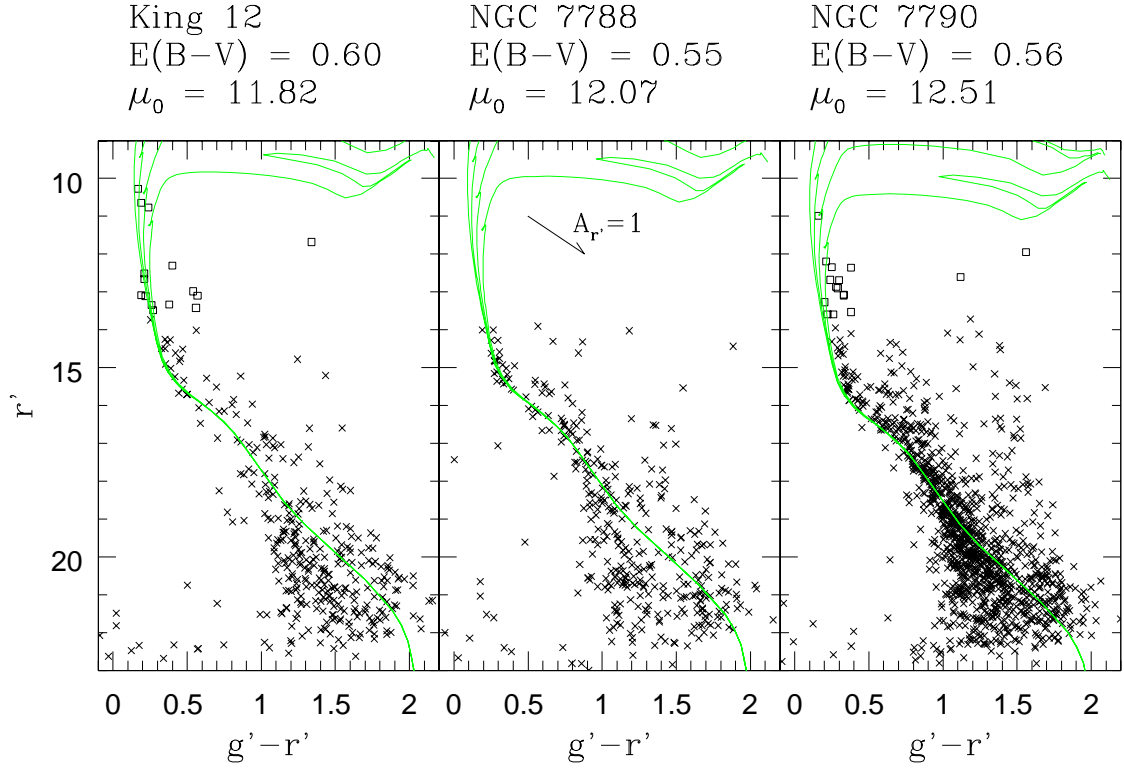}
\caption{The same as Figure 4, but showing Z = 0.019 
isochrones with reddenings and distance moduli that have been adjusted so that 
the isochrones better match the observations. The reddening and distance modulus of NGC 
7790 found from these data are not significantly different from other recently published 
values.}
\end{figure}

\clearpage

\begin{figure}
\figurenum{6}
\epsscale{1.00}
\plotone{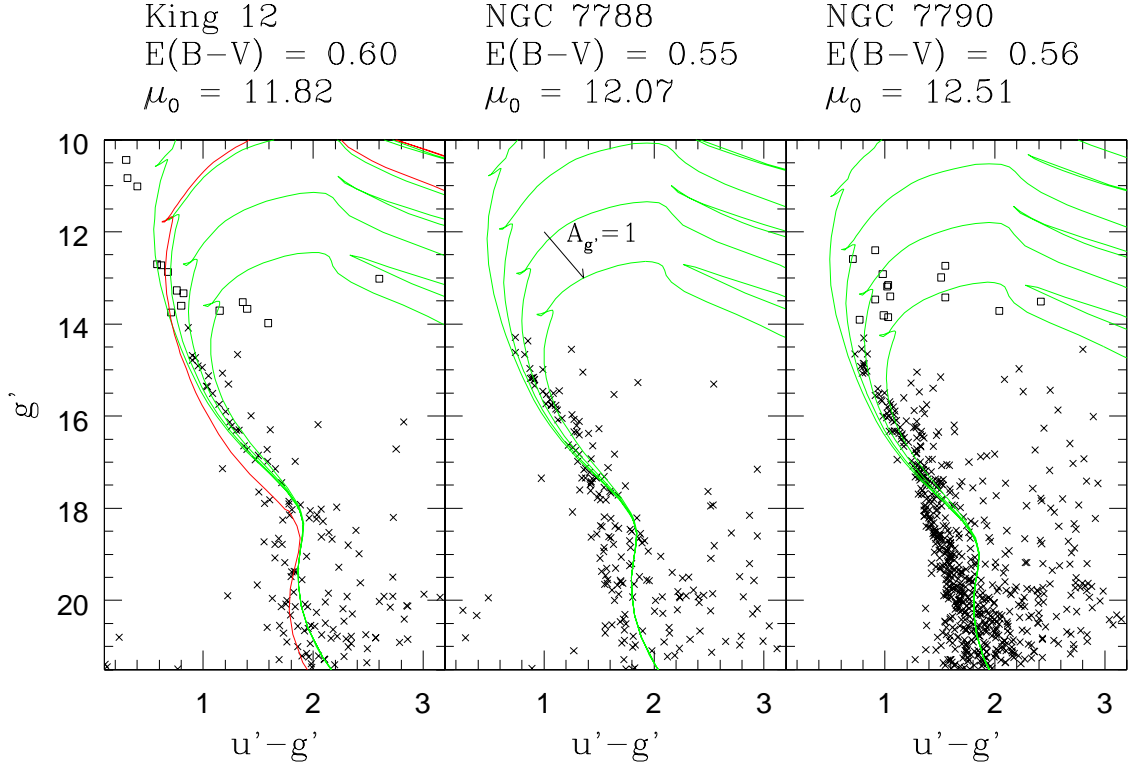}
\caption{The $(g', u'-g')$ CMDs of the clusters. The green lines are 
Z = 0.019 isochrones with ages of 10, 20, 40, and 80 Myr from the Girardi et al. 
(2004) compilation, while the red line shows a 20 Myr sequence with Z = 0.008. 
A reddening vector, with a length that corresponds to A${_g'} = 1$ mag, is 
shown in the NGC 7788 panel.}
\end{figure}

\clearpage

\begin{figure}
\figurenum{7}
\epsscale{1.00}
\plotone{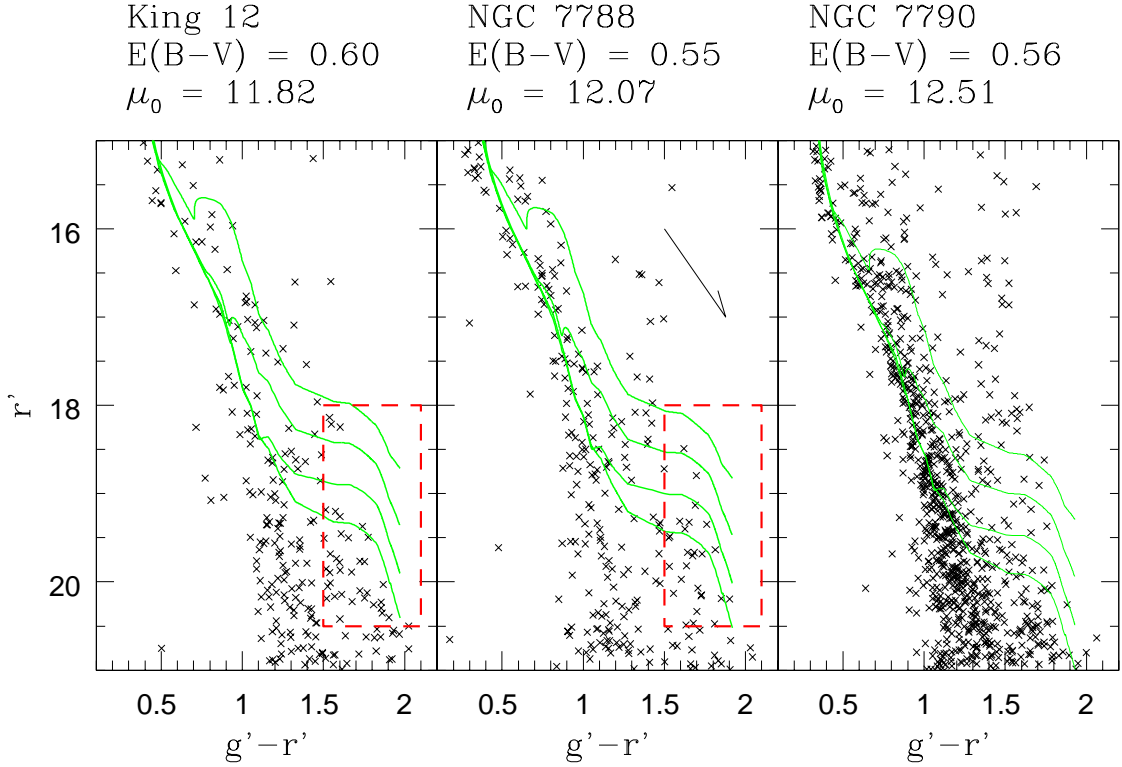}
\caption{The $(r', g'-r')$ CMDs of King 12, NGC 7788, and NGC 7790 are compared with 
Z=0.015 PMS isochrones, transformed from the Tognelli et al. (2011) models. 
The isochrones have ages of 10 Myr, 20 Myr, 40 Myr, and 80 Myr, and assume 
$Y=0.278$, $\alpha = 1.68$, and $X_D = 2 \times 10^{-5}$. 
The models have been transformed onto the observational plane using calibrations 
based on main sequence stars, with no allowance for circumstellar extinction. 
A reddening vector, with a length that corresponds to A${_g'} = 1$, is 
shown in the NGC 7788 panel. The red dashed boxes indicate the regions of the King 12 and 
NGC 7788 CMDs that contain an excess number of objects when compared with 
the surrounding field. The PMS sequences pass through this part of the CMD, suggesting 
that the excess population consists of PMS stars.}
\end{figure}

\clearpage

\begin{figure}
\figurenum{8}
\epsscale{0.75}
\plotone{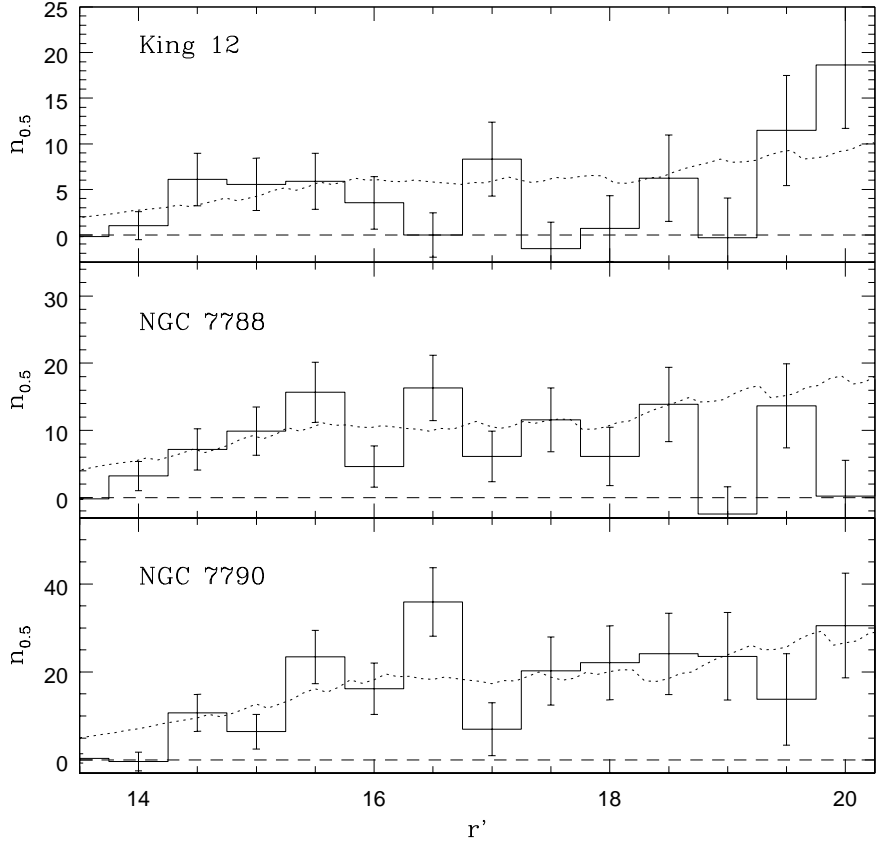}
\caption{The differential $r'$ LFs of King 12, NGC 7788 and NGC 7790. 
n$_{0.5}$ is the number of stars per 0.5 $r'$ magnitude interval, 
corrected for field star contamination. The error bars show 
$1\sigma$ random uncertainties computed from counting 
statistics. The dotted line is a model LF for a Z = 0.019 
population with an age 30 Myr and a Kroupa (1998) IMF. 
The model assumes that all stars form simultaneously on the main 
sequence, and so does not include the time it takes for stars to 
relax onto the main sequence. The LF of young clusters will then 
depart from this model at a magnitude that depends on their age. The model has been 
scaled to match the cluster LFs in the interval $r' = 14.5$ to 16.0. 
There is reasonable agreement between the model and the LFs of NGC 7788 
and NGC 7790 throughout much of the sampled magnitude range. 
At intermediate magnitudes there is a general tendency for the King 
12 counts to fall below the model predictions, and a similar break is also seen 
near the faint end of the NGC 7788 LF.}
\end{figure}

\clearpage

\begin{figure}
\figurenum{9}
\epsscale{0.75}
\plotone{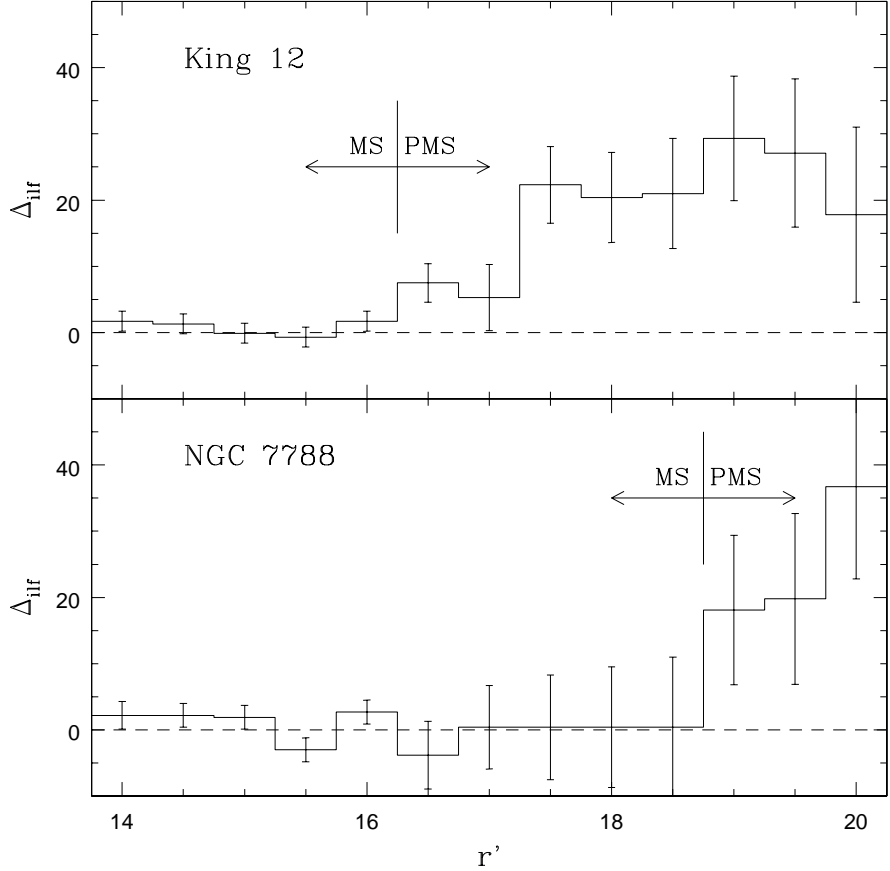}
\caption{The differences between the observed and model integrated LFs of King 12 and 
NGC 7788. The integrated LFs were calculated from the differential LFs in Figure 8. 
$\Delta_{ilf}$ is the difference between the integrated LFs per 0.5 magnitude $r'$ 
interval, in the sense model minus observed; positive values of $\Delta_{ilf}$ thus 
correspond to deficiencies in the number of stars in the observed LFs when 
compared with the models. The magnitudes that are adopted as the lower end of 
the main sequence and the onset of the PMS are indicated. The possibility that 
the onset of the PMS in King 12 may occur at $r' = 17.25$, where there is a large 
change in $\Delta_{ilf}$, rather than at $r' = 16.25$, is discussed in the text.}
\end{figure}

\clearpage

\begin{figure}
\figurenum{10}
\epsscale{0.75}
\plotone{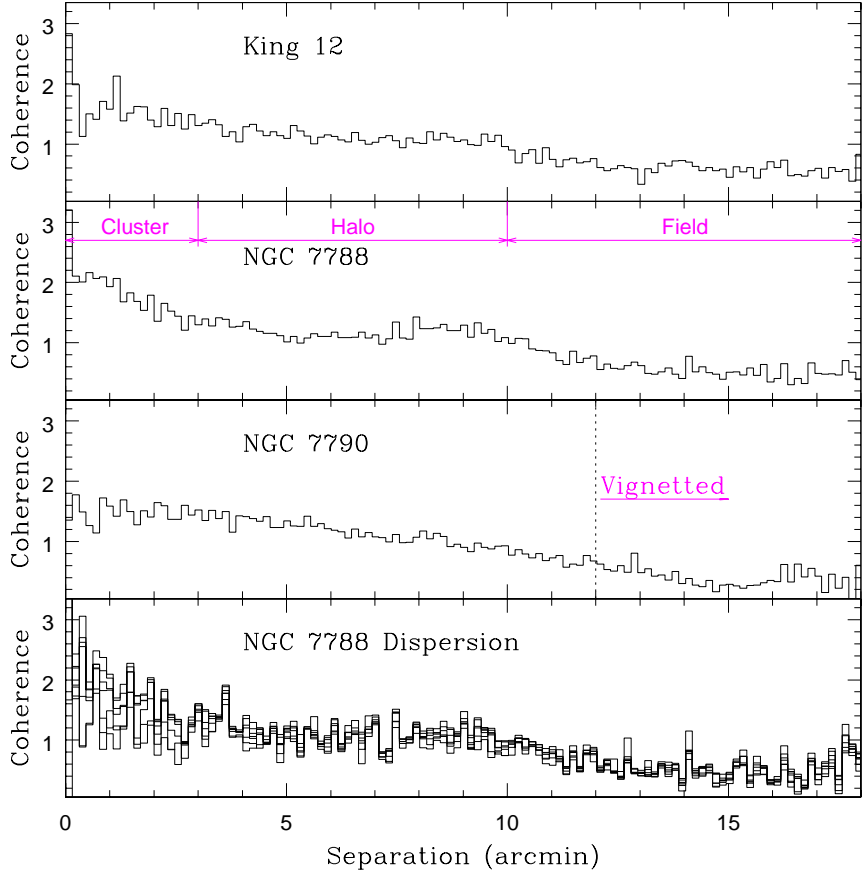}
\caption{The TPCFs obtained by ratioing the SFs of objects in the BMS and RPS samples 
within a 9 arcmin radius of King 12, NGC 7788, and NGC 7790. The curves 
have been normalized according to the number of sources in each sample. 
The edge of the MegaCam field is 6 arcmin from the center of NGC 7790, 
and so structural information from the southernmost regions of this cluster 
is missed. The structural regimes in the TPCFs of King 12 and NGC 
7788 that are discussed in the text are indicated in the NGC 7788 panel. 
The inflexion points in the TPCFs indicate that the main body of King 12 
extends out to separations of 2 arcmin (1 arcmin radius), while the main body of NGC 7788 
extends out to 3 arcmin separations (1.5 arcmin radius). There is a 
uniformly distributed population of BMS stars near King 12 and NGC 7788, which is 
interpreted as a halo around each cluster. The halo extends out to 10 arcmin separations 
(5 arcmin radii) in both clusters. BMS stars in NGC 7790 have a roughly uniform 
distribution out to separations of 4 -- 5 arcmin (2 -- 
2.5 arcmin radius), and the TPCF flattens at separations $> 15$ 
arcmin ($> 7.5$ arcmin radius). The projected size of NGC 7790 thus
exceeds those of King 12 and NGC 7788, even including their halos. The lower panel 
shows the dispersion in the NGC 7788 TPCF that results from jackknife re-sampling 
experiments, in which TPCFs were calculated after extracting 25\% of the points in 
the BMS and RPS samples at random.}
\end{figure}

\clearpage

\begin{figure}
\figurenum{11}
\epsscale{0.75}
\plotone{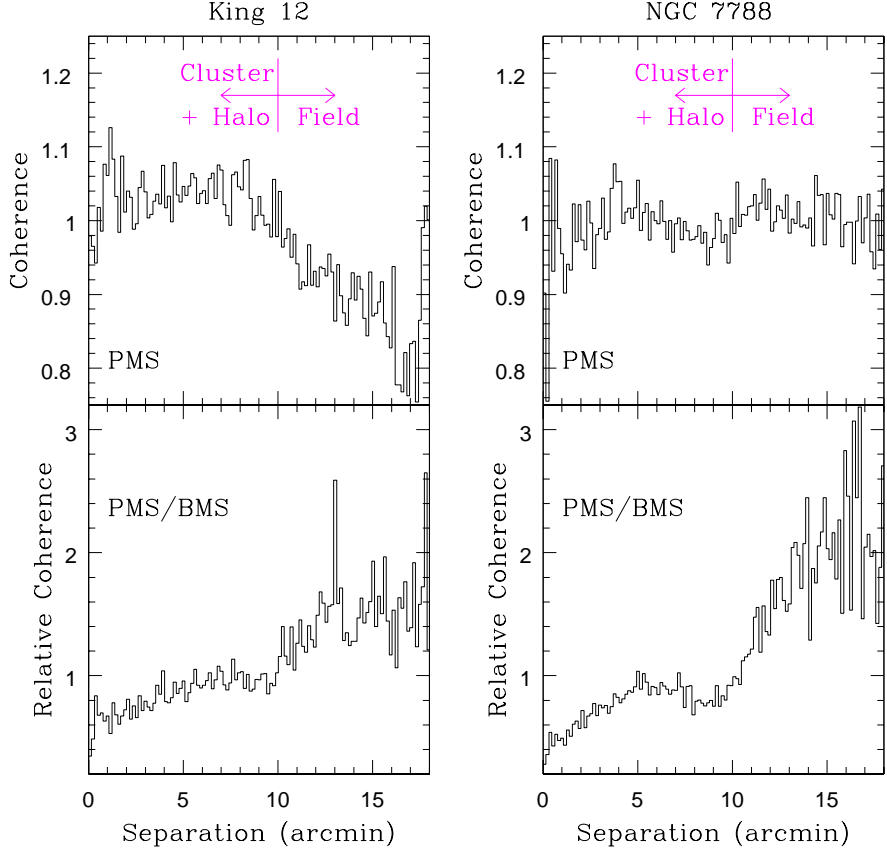}
\caption{The clustering properties of objects in the King 12 
and NGC 7788 PMS samples are investigated in this figure. (Top row) The TPCFs 
of possible PMS objects in King 12 and NGC 7788, obtained by ratioing the 
SF of PMS and RPS sources in each cluster, and then normalizing 
according to the number of objects in each sample. PMS sources in King 12 
show the highest degree of coherence at separations $< 10$ arcmin, in agreement with 
the clustering properties of BMS stars in this cluster. That PMS stars are detected in 
the halo regions of King 12, along with bright main sequence 
stars, indicates that the outer region of this cluster is 
populated by objects with a range of masses. In contrast to King 12, the TPCF of PMS 
objects near NGC 7788 is indicative of a uniformly distributed population, with 
no evidence of clustering. (Bottom row) The ratio of the PMS and BMS SFs in each 
cluster, normalized according to the number of objects in each sample. 
The clustering properties of BMS and PMS stars differ in both King 12 and NGC 
7788. This may reflect -- at least in part -- differing degrees of contamination from 
non-cluster members in the very different parts of the CMDs occupied by these samples.}
\end{figure}

\end{document}